\documentclass[aps,prd,twocolumn,superscriptaddress,tightenlines,nofootinbib]{revtex4-1}

\usepackage{amsfonts,amsmath,amsthm,amssymb}
\usepackage{mathrsfs}
\usepackage{bm}
\usepackage{color}
\usepackage{epsfig}
\usepackage{bigints}

\usepackage[usenames,dvipsnames]{xcolor}
\definecolor{orange}{cmyk}{0,0.5,1,0}
\definecolor{rossoCP3}{cmyk}{0,.88,.77,.40}
\definecolor{graa}{rgb}{0.8,0.8,0.8}
\definecolor{blaa}{rgb}{0.2,0.2,0.6}

\newcommand{\PRE}[1]{{#1}}   

\newcommand{\comment}[1]{}

\newcommand{\gsim}{\mathrel{\vcenter{\hbox{$>$}\nointerlineskip\hbox{$\sim$}}}}

\newcommand{\half}{\frac{1}{2}}

\newcommand{\Eps}{\mathcal{E}}

\newcommand{\Rearth}{R_\oplus}
\newcommand{\beq}[1]{\begin{equation}\label{#1}}
\newcommand{\eeq}{\end{equation}}
\newcommand\be{\begin{equation}}
\newcommand\ee{\end{equation}}
\newcommand{\bea}[1]{\begin{eqnarray} \label{#1}}
\newcommand{\eea}{\end{eqnarray}}
\newcommand{\ba}{\begin{eqnarray}}
\newcommand{\ea}{\end{eqnarray}}
\newcommand{\barr}{\begin{array}}
\newcommand{\earr}{\end{array}}

\begin{document}

\title{ 
 \PRE{\vspace*{0.9in}} \color{rossoCP3}{ 
Upgoing ANITA events as evidence of the CPT symmetric universe
}}

\author{Luis A. Anchordoqui}

\affiliation{Department of Physics \& Astronomy, Lehman College,  City University
  of New York, NY 10468, USA}
\affiliation{Department of Physics,
 Graduate Center, City University
  of New York,  NY 10016, USA}
\affiliation{Department of Astrophysics,
 American Museum of Natural History, NY
 10024, USA}

\author{Vernon Barger}
\affiliation{Department of Physics, University of Wisconsin, Madison, WI 53706, USA}

\author{John G.~Learned}
\affiliation{Department of Physics \& Astronomy, University of Hawaii at Manoa, Honolulu, HI 96822, USA}

\author{Danny Marfatia}
\affiliation{Department of Physics \& Astronomy, University of Hawaii at Manoa, Honolulu, HI 96822, USA}

\author{Thomas J. Weiler}
\affiliation{Department of Physics \& Astronomy, Vanderbilt University, Nashville TN 37235, USA}


\begin{abstract}
\noindent 
We explain the two upgoing ultra-high energy shower events observed by
ANITA as arising from the decay in the Earth's interior of the quasi-stable dark matter
candidate in the CPT symmetric universe. The dark matter particle is a
480~PeV right-handed neutrino that decays into a Higgs boson and a light
Majorana neutrino. The latter interacts in the Earth's crust to
produce a $\tau$ lepton that in turn initiates an atmospheric upgoing shower. The fact that
both events emerge at the same angle from the Antarctic ice-cap suggests an atypical dark matter density distribution in the Earth.
\end{abstract}

%
\maketitle

The three balloon flights of the ANITA experiment have resulted in the
observation of two unusual upgoing showers with energies of ($600 \pm
400)~{\rm PeV}$~\cite{Gorham:2016zah} and
($560^{+300}_{-200}$)~PeV~\cite{Gorham:2018ydl}. The energy estimates
are made under the assumption that the showers are initiated close to
the event's projected position on the ice. These estimates are lowered
significantly if the showers are initiated far above the ice. For
example, the energy of the second event is lowered by 30\% if the
shower is initiated four kilometers above the
ice~\cite{Gorham:2018ydl}.  In principle, these events could originate
in the atmospheric decay of an upgoing $\tau$-lepton produced through
a charged current interaction of $\nu_\tau$ inside the Earth. However,
the relatively steep arrival angles of these events ($27.4^\circ$ and $35^\circ$ above the horizon) 
create a tension
with the standard model (SM) neutrino-nucleon interaction cross
section. In particular, the second event implies a propagating
chord distance through the Earth of roughly $7.2 \times 10^3~{\rm
  km}$, which corresponds to $1.9 \times 10^4~{\rm km}$ water
equivalent (w.e.)
and a total of 18 SM interaction lengths at $E_\nu \sim 10^3~{\rm
  PeV}$.\footnote{The first event emerged at an angle of $27.4^\circ$
  above the horizon, implying a chord through the Earth of $5.5 \times
  10^3~{\rm km}$, which corresponds to $1.5 \times 10^4~{\rm km}$ w. e.
  for Earth's density profile~\cite{Gorham:2016zah}.
} Noting that the energy deposited in a shower is
roughly 80\% of the incident neutrino energy, our cosmic neutrino
energy range of interest is $200 \alt E_\nu/{\rm PeV} \alt 1000$. At
these energies, the neutrino flux is attenuated by a factor of
$10^{8}$~\cite{Alvarez-Muniz:2017mpk}. The ANITA Collaboration
concluded that a strong transient flux from a source with a compact
angular extent is required to avoid exceeding current bounds on
diffuse, isotropic neutrino fluxes~\cite{Gorham:2018ydl}. In this
Letter we provide an alternative mechanism that produces ${\cal O}
(100~{\rm PeV})$ $\tau$ leptons that exit the Earth's crust.

Neither cosmic ray observatories nor the IceCube telescope have seen
any anomalies at comparable energies. So we start with a discussion of
how the observation of the anomalous upgoing events at ANITA is
consistent with the non-observation of similar events at cosmic ray
facilities and IceCube.

Cosmic ray facilities have seen downgoing shower events with energies
up to $\sim 10^{5}~{\rm PeV}$, but have not reported any anomalous
upgoing showers~\cite{Patrignani:2016xqp}.  The IceCube Collaboration
has not reported any events above
10~PeV~\cite{Aartsen:2014gkd,Aartsen:2016xlq}. However, it has been
suggested that an upgoing track event from $\sim 11.5^\circ$ below the
horizon, with a deposited energy of $(2.6 \pm 0.3)~{\rm PeV}$ and
estimated median muon energy of $(4.5 \pm 1.2)~{\rm
  PeV}$~\cite{Aartsen:2016xlq}, could arise from an ${\cal
  O}(100)$~PeV $\tau$ lepton~\cite{Kistler:2016ask}.

ANITA measures the radio emission from the secondary electromagnetic
cascade induced by a neutrino interaction within the Antarctic ice
sheet. At a float altitude of 35~km, ANITA has a viewing area of
$10^6~{\rm km}^2$~\cite{Hoover:2010qt}.  Cosmic ray facilities have
viewing areas that are small compared to that of ANITA.  However,
transmission losses through the ice and beam efficiency at the
detector reduce the average acceptance solid angle of ANITA near the
horizon to $3.8 \times 10^{-4}~{\rm km^2 \ sr}$ at $10~{\rm
  PeV}$~\cite{Allison:2018cxu}. Moreover, some cosmic ray experiments
have been collecting data for more than 10~years, whereas ANITA has
collected data over three balloon flights
to yield a total live time of 53~days~\cite{Gorham:2018ydl,Anitalive}. Consequently, the
exposures of cosmic ray facilities to detect SM neutrino interactions
near the horizon exceed that of ANITA by about a factor of 60~\cite{Romero-Wolf}. Hence,
SM neutrino event rates at these experiments should exceed that of
ANITA. We may conclude that an explanation of the unusual
ANITA events that depends on an extraterrestrial isotropic flux of
high-energy $\nu_\tau$'s producing $\tau$ leptons that decay in the
atmosphere is highly disfavored. Leaving aside fine-tuned anisotropic
$\nu_\tau$ fluxes, we also conclude that the exotic ANITA signal
must originate inside the Earth.  Ground-based cosmic ray facilities
only search for quasi-horizontal air showers produced by
Earth-skimming neutrinos, i.e., those that are incoming at a few
degrees below the horizon~\cite{Feng:2001ue}. Therefore, if the
anomalous events originating inside the Earth are only visible at
large angles below the horizon, they escape detection at cosmic
ray facilities. Cosmic ray fluorescence detectors are sensitive to
  upgoing showers emerging at large angles above the horizon, but they 
  operate with a 10\% duty cycle.

IceCube looks for shower and track events in their cubic
kilometer under-ice laboratory. For showers emerging
  at $\sim 35^\circ$ above the horizon, the $\sim 1~{\rm km}^2$
  geometric area of IceCube is comparable to ANITA's effective area of
  $\sim 4~{\rm km}^2$~\cite{Gorham:2018ydl}. Then, a comparison of the expected number
  of events at IceCube and ANITA follows from the product of
  their geometric volumes and their live times~\cite{Gorham:2018ydl,Anitalive,IClive}:
\begin{equation}
{{\#\ \rm IceCube\ events}\over {\#\ \rm  ANITA\ events}} \sim \frac{1~{\rm km}^3 \times 2078~{\rm day}}{4~{\rm km}^2  \times
  {\rm depth}\times  53~{\rm day}} \simeq \frac{10~{\rm km}}{{\rm depth} }\,.\nonumber
\end{equation}
The range of depths at which the shower of an ANITA event is initiated
determines the the uncertainty in its energy.  It is possible that the
second event was initiated between an ice-depth of 3.22~km and a
height of 4~km above the ice~\cite{Gorham:2018ydl}.  We may then
expect a comparable number of events at IceCube and ANITA. If the
typical depth of shower initiation for ANITA is taken to be 4~km, then
IceCube should have seen 5 events.  As mentioned above, the 2.6~PeV
IceCube event may have its origin in an ${\cal O}(100)$~PeV $\tau$
lepton. Since the 95\% confidence level interval for observing 1 event
with no expected background is $[0.05,5.14]$~\cite{Feldman}, IceCube
data may not be in tension with ANITA's 2 events.

  It is compelling that the two ANITA
events are similar in energy and were observed at roughly the same
angle above the horizon. We speculate that these two events have
similar energies because they result from the two-body decay of a new
quasi-stable relic, itself gravitationally trapped inside the
Earth. (An alternative new physics interpretation considers a
  sterile neutrino propagating through the Earth which could scatter with
  nucleons via mixing to produce a $\tau$ lepton~\cite{Cherry:2018rxj}.)

  We frame our discussion in the context of the CPT symmetric
  universe~\cite{Boyle:2018tzc,Boyle:2018rgh}. In this scenario the
  universe before the Big Bang and the universe after the Big Bang is
  reinterpreted as a universe/anti-universe pair that is created from
  nothing. If the matter fields are described by the minimal extension
  of the SM with 3 right-handed neutrinos, then the only possible dark
  matter candidate is one of the right-handed neutrinos, say
  $\nu_{R,1}$. For this neutrino to be exactly stable the SM couplings
  must respect the $\mathbb{Z}_2$ symmetry, $\nu_{R,1} \to -
  \nu_{R,1}$. In the limit in which $\nu_{R,1}$ becomes stable, it
  also decouples from SM particles, i.e., $\nu_{R,1}$ only interacts
  via gravity.  To accommodate the present-day dark matter density,
  $\rho_{\rm DM} \approx 9.7 \times10^{-48}~{\rm GeV}^4$, the
  quasi-stable right-handed neutrino must have a mass $M \approx
  480~{\rm PeV}$~\cite{Boyle:2018tzc,Boyle:2018rgh}. Another relevant
  prediction of the CPT symmetric universe is that the three active
  neutrinos are Majorana particles as they obtain their masses by the
  usual seesaw mechanism.

Herein we assume that the $\mathbb Z_2$ symmetry is only
approximate. Note that in principle the non-gravitational couplings of
$\nu_{R,1}$ do not have to vanish, but have to be small enough so that
$\nu_{R,1}$ has a lifetime $\tau_{\nu_{R,1}} \gg H_0^{-1} = 9.778 \,
h^{-1}~{\rm Gyr}$, where $h \sim 0.68$. This opens up the possibility
to indirectly observe $\nu_{R,1}$ through its decay products. For
two-body decays, conservation of angular momentum forces the
$\nu_{R,1}$ to decay into a Higgs boson and a light Majorana neutrino.
The non-observation of a monochromatic neutrino signal from the
Galactic center or the Galactic halo sets a lower bound on the
lifetime of the quasi-stable right-handed neutrino, $\tau_{\nu_{R,1}}
\gsim 10^{28}~{\rm s}$~\cite{Aisati:2015vma,Rott:2017mxp}.  The decay
of the Higgs to $b \bar b$ results in a photon flux that is
constrained by gamma-ray data.  With an appropriate rescaling of
energy, the results of Ref.~\cite{Kuznetsov:2016fjt} show that the
gamma-ray constraint is more than 7 orders of magnitude weaker than
the neutrino line constraint.

A dense population of $\nu_{R,1}$ is expected at the center of the
Earth because as the Earth moves through the halo, the $\nu_{R,1}$
scatter with Earth matter, lose energy and become gravitationally
trapped. An accumulated $\nu_{R,1}$ then decays into a Higgs and an
active neutrino that propagates through the Earth and produces a
$\tau$ lepton near the Earth's surface.  The particular angle of the
ANITA events is a combination of the dark matter distribution in the
Earth, the neutrino interaction cross section, and the $\tau$ survival
probability.  The non-gravitational couplings have to be chosen to
produce a long lifetime and the needed abundance of right-handed
neutrinos in the Earth to yield the two ANITA events. To achieve a
sizable dark matter density in the Earth self-interactions may be
invoked.

\begin{figure}
\centering
\includegraphics[width=0.42\textwidth]{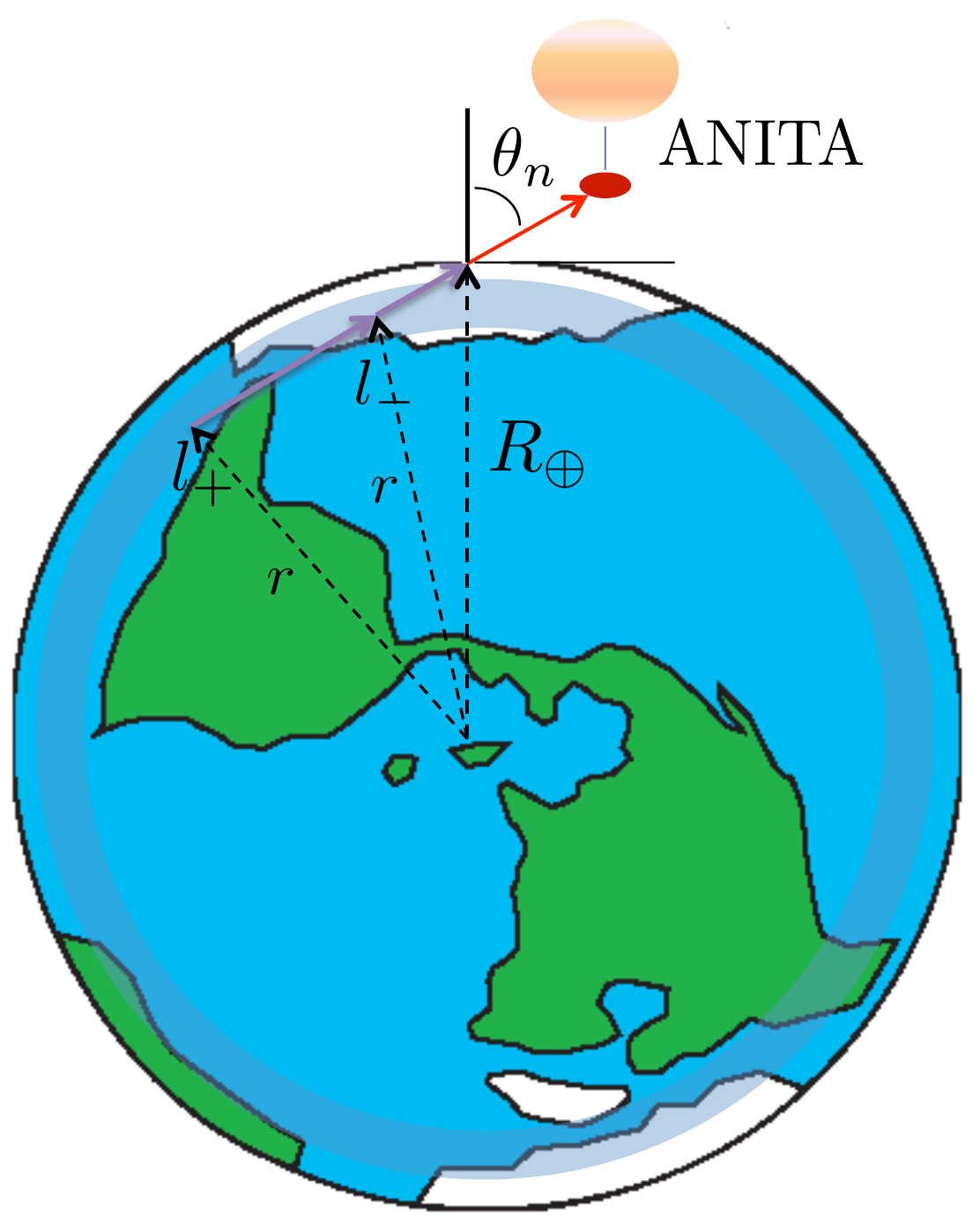}
\caption{The particle's trajectory to ANITA at a given nadir angle.
}
\label{fig}
\end{figure}

The event rate integrated over the entire Earth at a particular time  is 
\begin{equation}
{\rm Rate} \equiv \frac{dN}{dt} = 4\pi\int_0^{\Rearth} r^2\,dr\ \frac{n(r,t)}{\tau_{\nu_{R,1}}}\,,\nonumber
\label{wholeEarth}
\end{equation}
where $n(r,t)$ is the number density of $\nu_{R,1}$ at time $t$ and
$R_\oplus$ is the Earth's radius.  The observable rate
today ($t=t_0$), as a function of nadir angle $\theta_n$ 
is given by
\begin{eqnarray}
A_{\rm eff}\frac{d\,{\rm Rate}}{d\, |\cos\theta_n|} & = & 2\pi A_0\times 2\pi \int^{\Rearth}_{ {\Rearth} {\sin\theta_n} }\,
r^2 dr \, \frac{n(r,t_0)}{\tau_{\nu_{R,1}}}  \nonumber \\
& \times &  \left( e^{-(l_+/\lambda)} + e^{-(l_-/\lambda)} \right) \,
\Eps(\theta_n)\,,
\label{two}
\end{eqnarray}
where $l_\pm$ are the roots of  $\Rearth^2+l^2-2\Rearth l \cos\theta_n = r^2$,
i.e., 
\begin{equation}
l_\pm = \Rearth \left(\cos\theta_n \pm \sqrt{ \left(
      \frac{r}{\Rearth} \right)^2 -\sin^2 \theta_n } \ \right) \, ,\nonumber
\end{equation}
and $\lambda = 1.7 \times 10^7/(\sigma/{\rm
  pb}) ~{\rm km \, w.e.}$ is the mean-free-path, with $\sigma$ the neutrino-nucleon charged-current cross section. Here, the effective area $A_{\rm eff}=A_0\Eps(\theta_n)$ defines
the experimental efficiency $\Eps$ that includes the target area dependence on $\theta_n$ but 
not the $e^{-l/\lambda}$ suppression
which is given explicitly in the integrand. Note that $\Eps(\theta_n)$ vanishes for
$\theta_n<35^\circ$, peaks at about $75^\circ$, and vanishes above $85^\circ$~\cite{private}. In Eq.~(\ref{two}) we have neglected energy losses due to
neutral current interactions and effects from $\nu_\tau$
regeneration~\cite{Halzen:1998be}. For $200 \alt E_\nu/{\rm PeV} \alt
1000$, these effects are not important. For a 100~PeV neutrino,
$\sigma \sim 4.43 \times 10^3~{\rm pb}$, the interaction length in
rock is $\lambda \sim 10^3~{\rm km}$, and the average range of the
outgoing $\tau$ lepton is a few
km~\cite{Gandhi:1998ri,Dutta:2005yt}. Integrating over the
duration of an experiment yields the event number as opposed to
the event rate.

The fact that for fixed $r$, we have two special values of $l$,
i.e., $l_\pm$, can be seen from Fig.~\ref{fig}. Of course, if $r$ is too small, then the trajectory at
fixed $\theta_n$ does not intersect the circle at
all; this is the origin of the lower limit in the integration over $dr$. 

The exponential suppression factor in Eq.~(\ref{two}) can be written as 
\begin{eqnarray}
e^{-(l_+/\lambda)} + e^{-(l_-/\lambda)}& = & 2
\exp\left( - \frac{\Rearth\,\cos\theta_n}{\lambda}  \right) \nonumber \\
& \times & 
\cosh \left(\frac{\sqrt{r^2-\Rearth ^2\sin^2\theta_n}}{\lambda}
\right).
\label{cosh}
\end{eqnarray}
The competition between the falling (with increasing $\theta_n$)
$e^{-\Rearth\cos\theta_n/\lambda}$ term and the rising
$\Eps(\theta_n)$ term in Eq.~(\ref{two}) determines the most probable
angle of observation.  The two unusual ANITA events occur at
$27.4^\circ$ and $35^\circ$ above the horizon, so we may set the peak
of the distribution at $\sim 30^\circ$ above the horizon,
corresponding to a nadir angle of $\theta_n \sim 60^\circ$. So, taking
the view that the event distribution is maximized at
$\theta_n=60^\circ$ by a combination of ANITA's efficiency and the
dark matter distribution in the Earth, we require
\begin{equation}
\left. \frac{d^2\,{\rm Rate}}{d\, |\cos\theta_n|^2} \right|_{\cos \theta_n
  = \half}  = 0  \,.
\label{condition}
\end{equation} 
This result becomes a constraint on the model parameters in Eq.~(\ref{two}). 

We end with three observations: {\it (i)}~It is generally assumed
that after the dark matter particles become gravitationally bound,
they quickly lose their momentum and sink to the core of Earth~\cite{Gould:1987ir}. We have proposed that ANITA data may be
indicating that the dark matter distribution in the Earth may be more complicated.
This may result from a recent encounter of the Earth with a dark
disk.\footnote{Cosmological N-body
simulations suggest that a thick dark disk is formed
naturally in Milky Way-type galaxies as a consequence of satellite
mergers (which usually get dragged into the plane of their host galaxy~\cite{Read:2008fh}. This paradigm is consistent with observations~\cite{Kramer:2016dew}.} {\it (ii)}~Quasi-stable right-handed neutrinos will also
accumulate in the core of the Sun and the Moon, and on decay will produce a flux of high-energy
neutrinos. However, the neutrinos will not escape the Sun or the
Moon, and the latter does not have an atmosphere in which the
$\tau$ leptons can produce showers, so consequently the flux from these
sources is unobservable.  {\it (iii)}~ Data from the fourth ANITA flight is currently being analyzed and may lead to further enlightenment.
The second generation of the
Extreme Universe Space Observatory (EUSO) instrument, to be flown
aboard a super-pressure balloon (SPB) in 2022 will monitor the night
sky of the Southern hemisphere for upgoing showers emerging at large
angles below the horizon~\cite{Adams:2017fjh}. EUSO-SPB2 will provide
an important test both of the unusual ANITA events and of the ideas
discussed in this Letter. \\

This work has been supported by the U.S. National Science Foundation
(NSF Grant PHY-1620661), the Department of Energy (DoE Grants
DE-SC0017647, DE-SC0010504, and DE-SC-0011981), and the National
Aeronautics and Space Administration (NASA Grant 80NSSC18K0464).
V.B. thanks KITP at UCSB for its hospitality while this work was in
progress.

\end{document}